\documentstyle[12pt]{article}

\begin{document}

\def\thefootnote{\fnsymbol{footnote}}

{\it University of Shizuoka}

\hspace*{10cm} {\bf Revised Version of}\\[-.3in]

\hspace*{10cm} {\bf US-97-01}\\[-.3in]

\hspace*{10cm} {\bf hep-ph/9701345}\\[-.3in]

\hspace*{10cm} {\bf May 1997}\\[.4in]

\vspace*{1cm}
\begin{center}

{\large\bf  NNI-Form Quark Mass Matrix }\\[.1in]

{\large\bf Expressed by the Observable Quantities}\\[.3in]

{\bf Yoshio Koide\footnote{
E-mail address: koide@u-shizuoka-ken.ac.jp}} \\[.1in]

Department of Physics, University of Shizuoka \\[.1in] 
52-1 Yada, Shizuoka 422, Japan \\[.5in]

{\large\bf Abstract}\\[.1in]
\end{center}

\begin{quotation}
It is pointed out that 
the phase convention of the CKM matrix $V$  affects 
texture analysis of the quark mass matrices $(M_u, M_d)$  
when we try to describe $(M_u, M_d)$ by the observable quantities 
(quark masses and CKM matrix parameters) only. 
This is demonstrated for a case of the non-Hermitian Fritzsch-type 
mass matrix $(\widetilde{M}_u, \ \widetilde{M}_d)$, which is a general 
expression of quark mass matrix $(M_u, M_d)$ and is described by twelve 
parameters. 
We find that we can always choose a phase convention of $V$ which yields 
$\widetilde{M}_{u32} = 0$, 
so that the remaining ten parameters in $(\widetilde{M}_u, \ 
\widetilde{M}_d)$ can completely be expressed by the ten observable 
quantities.
\end{quotation}

\vfill

\vglue.1in
\noindent


\newpage

It is well-known that in a three-family model, quark mass
matrices $(M_u, M_d)$ are, in general, described by 36 parameters, 
while the number of the observable quantities is ten, i.e., six quark 
masses and four Cabibbo-Kobayashi-Maskawa (CKM) [1] matrix 
parameters. 
Once we give the quark mass matrices $(M_u, M_d)$, we can uniquely 
obtain the masses $(D_u, D_d)$ and CKM matrix $V$ from 
$$
\begin{array}{l}
U_L^u M_u U_R^{u\dagger} = D_u \equiv {\rm diag}(m_u, m_c, m_t) \ , \\
U_L^d M_d U_R^{d\dagger} = D_d \equiv {\rm diag}(m_d, m_s, m_b) \ , \\
\end{array} 
\eqno(1)
$$
$$
V= U_L^u U_L^{d\dagger} \ ,  
\eqno(2)
$$
except for the degree of freedom of the rephasing of $V$ 
(phase convention of $V$).
Looking for a clue to unified understanding of the quarks and leptons,
so far, many mass matrix models have been proposed.

On the other hand, in order to obtain a hint on such a unified mass 
matrix model, it has been tried [2,3] to describe the mass matrices 
$(M_u, M_d)$ by the observable quantities $(D_u, D_d)$ and $V$, 
without including any additional parameters.
However, such the inverse procedure $(D_u, D_d; V) \rightarrow 
(M_u, M_d)$ is not unique.
Only when we choose a special quark basis, we can describe $(M_u, M_d)$ 
by the observable quantities $(D_u, D_d; V)$ only.
Hereafter, let us call such a quark basis as the ``minimal parameter" 
basis.
It is well-known [2] that the quark basis on which up-quark mass matrix 
$M_u$ is diagonal is an example of the minimal parameter basis:
$$
\begin{array}{l}
H_u \equiv  M_u M_u^\dagger = \widehat{H}_u \equiv D_u^2 \ , \\
H_d \equiv M_d M_d^\dagger = \widehat{H}_d \equiv V D_d^2 V^\dagger \ . 
\end{array}
\eqno(3)
$$
To know the explicit form of the mass matrixes $(M_u, M_d)$ in  a 
minimal parameter basis will offer a useful hint on the realistic 
model-building of quark mass matrix. 
At present, it is not so well-known what any other minimal parameter 
basis there is.

Recently, there has been considerable interest [4,5] in the non-Hermitian 
Fritzsch-type quark mass matrix, i.e., 
the quark mass matrix with nearest-neighbor interactions (NNI) form, 
because of its simple form and generality.
The model is described by twelve parameters as stated later.
In the present paper, it is pointed out that the demand of the minimal 
parameter basis for the NNI-form mass matrices 
$(\widetilde{M}_u, \widetilde{M}_d)$  uniquely leads to 
$(\widetilde{M}_u)_{32}=0$, so that $(\widetilde{M}_u, \widetilde{M}_d)$ 
is described by ten parameters. 
It should be noticed that $(\widetilde{M}_u)_{32}=0$ is obtained by 
choosing a special phase convention of $V$. 
This conclusion might be felt somewhat strange, because the degree of 
freedom of the phase convention of $V$ comes from that of the rephasing 
of the physical (mass-eigenstate) quark fields $(u_i, d_i)$ ($i=1,2,3$).
It is well-known that the rephasing of the physical quarks is 
independent of the quark mass matrix structure $(M_u, M_d)$, and cannot 
change the structure $(M_u, M_d)$. 
The purpose of the present paper is to point out that in the inverse 
procedure $(D_u, D_d; V) \rightarrow (M_u, M_D)$, 
the rephasing of $V$ affects the textures of $(M_u, M_d)$, 
so that we must take  notice of the phase convention of $V$, 
although the procedure $(M_u, M_d)\rightarrow (D_u, D_d; V) $ is unique 
(except for the phase convention of $V$). 
We demonstrate it for the NNI mass matrices 
$(\widetilde{M}_u, \widetilde{M}_d)$, and we discuss why the rephasing 
of $V$ affects the textures of $(\widetilde{M}_u, \widetilde{M}_d)$.
(We will find that the ``rephasing" of $V$ in (3) is tacitly 
accompanied with the ``re-basing" of the quark fields 
$(u_L^0, d_L^0)$ which are eigenstates of weak interactions.)
The notice which is given for the NNI mass matrix 
$(\widetilde{M}_u, \widetilde{M}_d)$ will also be useful 
for general texture analysis [6] of the mass matrices $(M_u, M_d)$.

The NNI mass matrix $\widetilde{M}_q$ has the following structure:
$$
\widetilde{M}_q = P(\delta^q_L)\widetilde{\overline{M}}_q P^\dagger 
(\delta_R^q) \  , \ \ (q=u, d) \  , 
\eqno(4)
$$
$$
P(\delta) = {\rm diag} (e^{i\delta_1}, e^{i\delta_2}, 
e^{i\delta_3}) \  , 
\eqno(5)
$$
$$
\widetilde{\overline{M}}_q = \left(\begin{array}{ccc}
0 & c_1^q & 0 \\
c_2^q & 0 & b_1^q \\
0 & b_2^q & a_q \\
\end{array} \right) \  , 
\eqno(6)
$$
where $a_q, \ b_1^q, \ b_2^q, \ c_1^q$ and $c_2^q$ are real 
parameters. 
(Hereafter, we denote a matrix $M$ which consists of real 
elements as $\overline{M}$.)
The form has first been suggested by Fritzsch [7], and 
then it has been pointed out by Branco, Lavoura and Mota [8] 
that any mass matrix ($M_u, \ M_d$) can, without losing generality, 
be transformed into a NNI form $(\widetilde{M}_u, \ \widetilde{M}_d)$. 
Harayama and Okamura [5] have given the exact expression of the 
NNI-form  mass matrix which is described in terms of the ten observable 
quantities and two ``implicit" parameters, 
and have exhibited some special cases of the texture zeros. 
In the present paper, we will point out that the two ``implicit" 
parameters in the expression by Harayama and Okamura play a role 
of fixing the phase convention of $V$.

In general, the ten observable quantities are obtained by 
diagonalizing the Hermitian matrix 
$$
H_q \equiv M_q M_q^\dagger 
\eqno(7)
$$
as follows:
$$
D_q^2 \equiv {\rm diag} (m_1^q, \ m_2^q, \ m_3^q) = 
U_q H_q U_q^\dagger \  . 
\eqno(8)
$$
For the mass matrix with NNI form, those are given by 
$$
D_q^2 = R_q \widetilde{\overline{H}}_q R_q^T  \ , 
\eqno(9)
$$
$$
V = R_u P(\delta_L^d - \delta_L^u) R_d^T  \ , 
\eqno(10)
$$
where $\widetilde{\overline{H}}_q = \widetilde{\overline{M}}_q 
\widetilde{\overline{M}}_q^T$, and $R_q$ are orthogonal 
transformation matrices. Of the three phases  $\delta_{Li}\equiv
\delta_{Li}^d - \delta_{Li}^u$, the observable quantities are only two. 
Therefore, the mass matrix with NNI form, $(\widetilde{M}_u, \ 
\widetilde{M}_d)$, has, in general, twelve parameters ($5 + 5 + 2$).

In the present paper, inversely, we try to express the NNI form 
$(\widetilde{M}_u, \ \widetilde{M}_d)$ by the observable 
quantities only. 
At present, we know rough values of nine observable quantities (six 
quark masses and three CKM parameters) except for $CP$ violating 
phase $\delta$ ($\delta$ is defined by $V_{ub} = |V_{ub}| e^{-i\delta}$ 
in the standard parametrization of $V$ [9]). 
Therefore, one of our interests is to see the $\delta$-dependency of 
each matrix element $\widetilde{M}_{qij}$. 
It is well known that the CKM matrix $V$ has the degree of 
freedom of rephasing, i.e., $V' = P(\alpha) VP^\dagger(\beta)$ is 
physically equivalent to $V$. 
In the present paper, we will conclude that  the textures of 
$(\widetilde{M}_u, \ \widetilde{M}_d)$ are dependent on the phase
convention of $V$, and, for example, we can always take a phase 
convention of $V$  which yields $\widetilde{M}_{u32} = 0$.
In such the phase convention, $\widetilde{\overline{M}}_u$ and 
$\widetilde{\overline{M}}_d$ have four and five parameters, respectively, 
in addition to 
a relative phase parameter $\widetilde{\phi}$ of $\widetilde{M}_u$ 
to $\widetilde{M}_d$, 
so that the ten parameters in $(\widetilde{M}_u, \ \widetilde{M}_d)$ 
are sufficient to fix the ten observable quantities. 
The magnitudes of $\widetilde{M}_{dij}$ can approximately be fixed 
by three down-quark masses $(m_d, \ m_s, \ m_b)$ and three 
CKM matrix parameters $(|V_{us}|, \ |V_{cb}|, \ |V_{ub}|)$, 
because they are insensitive to the value of $CP$ violating 
phase $\delta$.

{}From (4), we obtain 
$$
\widetilde{H}_q = P (\delta_L^q) \widetilde{\overline{H}}_q P(-\delta_L^q) \  , 
\eqno(11)
$$
$$
\widetilde{\overline{H}}_q = \widetilde{\overline{M}}_q 
\widetilde{\overline{M}}_q^\dagger 
= \left(\begin{array}{ccc}
c_1^{q2} & 0 & b_2^q c_1^q \\ 
0 & b_1^{q2} + c_2^{q2} & a_q b_1^q \\ 
b_2^q c_1^q & a_q b_1^q & a_q^2 + b_2^{q2} \\ 
\end{array} \right) \  . 
\eqno(12)
$$
Therefore, if we know a matrix $(\widetilde{H}_u, \ \widetilde{H}_d)$ 
in which $\widetilde{H}_{u12} = \widetilde{H}_{d12} = 0$, we can obtain 
the NNI form $(\widetilde{M}_u, \ \widetilde{M}_d)$ from the relations
$$
a = \frac{1}{\sqrt{\widetilde{\overline{H}}_{11}}} \sqrt{\widetilde{\overline{H}}_{11} 
\widetilde{\overline{H}}_{33} - \widetilde{\overline{H}}_{31}^2 } \  , 
$$
$$
b_1 = \frac{\widetilde{\overline{H}}_{23} \sqrt{\widetilde{\overline{H}}_{11}}}
{\sqrt{\widetilde{\overline{H}}_{11} \widetilde{\overline{H}}_{33} - 
\widetilde{\overline{H}}_{31}^2}} \  , \ \ b_2 = \frac{\widetilde{\overline{H}}_{23}}
{\sqrt{\widetilde{\overline{H}}_{11}}} \  , 
$$
$$
c_1 = \sqrt{\widetilde{\overline{H}}_{11}} \  , \ \ c_2 
= \sqrt{\frac{{\rm det} \widetilde{\overline{H}}}
{\widetilde{\overline{H}}_{11} 
\widetilde{\overline{H}}_{33} - \widetilde{\overline{H}}_{31}^2}} \  , 
\eqno(13)
$$
where, for simplicity, we have omitted the index $q$. 

On the other hand, if we choose a specific quark basis [2] in 
which $H_u$ is diagonal, the Hermitian matrix $(H_u, \ H_d) \equiv 
(\widehat{H}_u, \ \widehat{H}_d)$ is given by (3).
Since $(H_u, \ H_d)$ has the degree of freedom of the rephasing, 
each $\phi_{ij} \equiv {\rm arg}\widehat{H}_{dij}$ is not 
observable quantity, 
but 
$$
\widehat{\phi} \equiv \widehat{\phi}_{12} + \widehat{\phi}_{23} 
+ \widehat{\phi}_{31} \  . 
\eqno(14)
$$
The ten observable quantities are given by three parameters in 
$\widehat{\overline{H}}_u \equiv D_u^2$, six parameters in 
$\widehat{\overline{H}}_d$ and one phase parameter 
$\widehat{\phi}$. Therefore, our 
task is to find a unitary transformation matrix $U$ which satisfies 
$$
\begin{array}{l}
U \widehat{H}_u U^\dagger = \widetilde{H}_u \  , \\
U \widehat{H}_d U^\dagger = \widetilde{H}_d \  . \\
\end{array} \eqno(15)
$$
For four parameters in $U$, there are four conditions, i.e., 
$(U \widehat{H}_u U^\dagger)_{12} = (U \widehat{H}_d U^\dagger)_{12} = 0$, 
so that we can fix $U$ when we fix the phase convention of $V$. 
In other words, the matrix $U$ depends on the phase convention of $V$. 

Let us choose a phase convention of $V$ in which $\widehat{\phi}_{12} 
= \widehat{\phi}_{31} = 0$. 
Then $(\widetilde{H}_u, \ \widetilde{H}_d)$ is given by 
$$
\begin{array}{l}
\widetilde{H}_u = R \widehat{H}_u R^T \  , \\
\widetilde{H}_d = R \widehat{H}_d R^T \  , \\
\end{array} \eqno(16)
$$
where $R$ is an orthogonal matrix 
$$
R = \left(\begin{array}{ccc}
1 & 0 & 0 \\
0 & c & s \\
0 & -s & c \\
\end{array} \right) \  , 
\eqno(17)
$$
$c \equiv \cos\theta$ and $s \equiv \sin\theta$. The condition 
$\widehat{H}_{d12} = 0$ leads to 
$$
s/c = -\widehat{\overline{H}}_{d12}/ \widehat{\overline{H}}_{d31} \  . 
\eqno(18)
$$
Each element of $\widehat{H}_d$ is real, except for $\widehat{H}_{d23}$. 
The element $\widehat{H}_{d23}$ is given by 
$$
\widehat{H}_{d23} = \widehat{\overline{H}}_{d23} 
e^{i \widetilde{\phi}} \  , 
\eqno(19)
$$
where 
$$
\tan\widetilde{\phi} = \frac{(\widehat{\overline{H}}_{d31}^2 + 
\widehat{\overline{H}}_{d12}^2) \widehat{\overline{H}}_{d23} 
\sin\widehat{\phi}}
{(\widehat{\overline{H}}_{d31}^2 - \widehat{\overline{H}}_{d12}^2) 
\widehat{\overline{H}}_{d23} \cos\widehat{\phi} - 
(\widehat{\overline{H}}_{d33} 
- \widehat{\overline{H}}_{d22})\widehat{\overline{H}}_{d12} 
\widehat{\overline{H}}_{d31}} \  . 
\eqno(20)
$$
Therefore, $\widetilde{H}_d$ is given by 
$$
\widetilde{H}_d = \widetilde{P} \widetilde{\overline{H}}_d 
\widetilde{P}^\dagger \  , 
\eqno(21)
$$
where 
$$
\widetilde{P} = {\rm diag} (1, e^{i\widetilde{\phi}}, 1) \  , 
\eqno(22)
$$
and 
$$
\widetilde{\overline{H}}_{d11} = \widehat{\overline{H}}_{d11} \  , 
$$
$$
\widetilde{\overline{H}}_{d22} 
= \frac{\widehat{\overline{H}}_{d22} \widehat{\overline{H}}_{d31}^2 
+ \widehat{\overline{H}}_{d33} \widehat{\overline{H}}_{d12}^2 
- 2\widehat{\overline{H}}_{d12} \widehat{\overline{H}}_{d23} 
\widehat{\overline{H}}_{d31} \cos \widehat{\phi}}{\widehat{\overline{H}}_{d12}^2 
+ \widehat{\overline{H}}_{d31}^{2}} \  , 
$$
$$
\widetilde{\overline{H}}_{d33} = \frac{\widehat{\overline{H}}_{d22} 
\widehat{\overline{H}}_{d12}^2 + \widehat{\overline{H}}_{d33} 
\widehat{\overline{H}}_{d31}^2 + 2\widehat{\overline{H}}_{d12} 
\widehat{\overline{H}}_{d23} \widehat{\overline{H}}_{d31} \cos\widehat{\phi}}
{\widehat{\overline{H}}_{d12}^2 + \widehat{\overline{H}}_{d31}^2} \  , 
$$
$$
\widetilde{\overline{H}}_{d12} = 0 \  , \ \ \widetilde{\overline{H}}_{d31} = 
\sqrt{\widehat{\overline{H}}_{d12}^2 + \widehat{\overline{H}}_{d31}^2} \  , 
$$
$$
\widetilde{\overline{H}}_{d23} = \frac{1}
{\sqrt{\widetilde{\overline{H}}_{d11}}} 
\sqrt{\widetilde{\overline{H}}_{d11} \widetilde{\overline{H}}_{d22} 
\widetilde{\overline{H}}_{d33} - \widetilde{\overline{H}}_{d22} 
\widetilde{\overline{H}}_{d31}^2 - m_d^2 m_s^2 m_b^2} \  . 
\eqno(23)
$$
On the other hand, $\widetilde{H}_u = \widetilde{\overline{H}}_u$ is given by 
$$
\widetilde{\overline{H}}_{u11} = m_u^2 \  , \ \ 
\widetilde{\overline{H}}_{u22} = c^2m_c^2 + s^2m_t^2 \  , \ \ 
\widetilde{\overline{H}}_{u33} = s^2m_c^2 + c^2m_t^2 \  , 
$$
$$
\widetilde{\overline{H}}_{u12} = \widetilde{\overline{H}}_{u31} = 0 \  , 
\ \ \widetilde{\overline{H}}_{u23} = sc (m_t^2 - m_c^2) \  . 
\eqno(24)
$$
{}From (20)--(24) and (13), we can obtain the exact form of 
$(\widetilde{M}_u, \ \widetilde{M}_d)$ 
expressed by the observable quantities only. 

In order to obtain the approximate form of $(\widetilde{M}_u, \ 
\widetilde{M}_d)$, we use an approximate expression of the CKM matrix $V$, 
$$
V \simeq \left(\begin{array}{ccc}
1-\frac{1}{2}\lambda^2 & \lambda & \sigma e^{-i\delta} \\ 
-\lambda & 1-\frac{1}{2}\lambda^2 & \rho \\ 
\lambda\rho - \sigma e^{i\delta} & -\rho & 1-\frac{1}{2}\rho^2 \\
\end{array} \right) \  , 
\eqno(25)
$$
where $\lambda = |V_{us}|, \ \rho=|V_{cb}|$ and $\sigma = |V_{ub}|$. 
Then, we obtain 
$$
\begin{array}{l}
(V D_d^2 V^\dagger)_{11} \simeq \sigma^2 m_b^2 (1 + x^2) \  , \\
(V D_d^2 V^\dagger)_{22} \simeq \rho^2m_b^2 (1+y^2/x^2) \  , \\
(V D_d^2 V^\dagger)_{33} \simeq m_b^2 \  , \\
(V D_d^2 V^\dagger)_{12} \simeq \rho \sigma m_b^2(e^{-i\delta} + y)\ , \\
(V D_d^2 V^\dagger)_{23} \simeq \rho m_b^2 \  , \\
(V D_d^2 V^\dagger)_{31} \simeq \sigma m_b^2 e^{i\delta} \  , \\
\end{array} \eqno(26)
$$
where 
$$
x = \frac{\lambda}{\sigma}\frac{m_s}{m_b} \  , \ \ 
y = \frac{\lambda}{\rho\sigma}\frac{m_s^2}{m_b^2}=
\frac{x}{\rho}\frac{m_s}{m_b} \ , 
\eqno(27)
$$
and we have used the observed relations $\rho \sim O(\lambda^2)$,  
$\sigma \sim O(\lambda^4)$,  $(m_s/m_b)^2 \sim O(\lambda^5)$ and 
$(m_d/m_b)^2 \sim O(\lambda^9)$. 
Since $\widehat{\phi}_{12} \simeq \tan^{-1}[-\sin\delta/(y+\cos\delta)]$, 
$\widehat{\phi}_{23} \simeq 0$ and $\widehat{\phi}_{31} \simeq \delta$, 
we obtain 
$$
\sin\widehat{\phi} \simeq 
\frac{y\sin\delta}{\sqrt{1 + y^2 +2y\cos\delta}} \  . 
\eqno(28)
$$
The parameter $s = \sin\theta$ in $R$ is given by 
$$
s \equiv \sin\theta \simeq -\rho \sqrt{1 + y^2 + 2y \cos \delta} \  . 
\eqno(29)
$$
Then, $\widetilde{\overline{H}}_{dij}$ are given by 
$$
\widetilde{\overline{H}}_{d11} \simeq \sigma^2 m_b^2 (1 + x^2) \  , \ \ 
\widetilde{\overline{H}}_{d22} \simeq x^2 m_s^2 \  , \ \ 
\widetilde{\overline{H}}_{d33} \simeq m_b^2 \  . 
$$
$$
\widetilde{\overline{H}}_{d12} = 0 \  , \ \ 
\widetilde{\overline{H}}_{d23} \simeq x^2 m_s m_b/\sqrt{1 + x^2} \  , \ \ 
\widetilde{\overline{H}}_{d31} \simeq \sigma m_b^2 \  , 
\eqno(30)
$$
so that we obtain 
$$
\widetilde{\overline{M}}_u \simeq \left(\begin{array}{ccc}
0 & m_u & 0 \\
m_c & 0 & s m_t \\
0 & 0 & c m_t \\
\end{array} \right) \  , 
\eqno(31)
$$
\renewcommand{\arraystretch}{2}
$$
\widetilde{\overline{M}}_d \simeq \left(\begin{array}{ccc}
0 & \displaystyle\frac{\sqrt{1 + x^2}}{x} \lambda m_s & 0 \\
\frac{1}{\lambda} m_d & 0 & x m_s \\
0 & \sqrt{1 + x^2} m_b & \displaystyle\frac{x}{\sqrt{1 + x^2}} m_b \\
\end{array} \right) \  . 
\eqno(32)
$$
\renewcommand{\arraystretch}{1}
Note that every value of $\widetilde{\overline{H}}_{dij}$ (and 
$\widetilde{\overline{M}}_{dij}$) is insensitive to 
the value of $\delta$. 
The results (31) and (32), of course, satisfy the 
relations $a_u c_1^u c_2^u = m_u m_c m_t $ and 
$a_d c_1^d c_2^d = m_d m_s m_b$, respectively, which come from 
${\rm det}\widetilde{H}_q = ({\rm det}\widetilde{M}_q)^2 
= (a_q c_1^q c_2^q)^2 = 
(m_1^q m_2^q m_3^q)^2$. 
Also note that the expressions (30) and (32) are valid in the 
the phase convention of $V$ in which ${\rm arg}(V D_d^2 V^\dagger)_{12}
={\rm arg}(V D_d^2 V^\dagger)_{31} = 0$.
Of course, the expression given by (31) and (32) is one of 
the solutions exhibited by Harayama and Okamura [5].

Thus, we obtain the NNI form 
$$
(\widetilde{M}_u, \ \widetilde{M}_d) = (P(\delta_L) \widetilde{\overline{M}}_u 
P^\dagger(\delta_R), \ P(\delta_L) \widetilde{P} \widetilde{\overline{M}}_d 
P^\dagger(\delta_R)) \  , 
\eqno(33)
$$
where the phase matrix $\widetilde{P}$ is 
defined by (22) with $\widetilde{\phi}$ given by 
$$
\tan \widetilde{\phi} \simeq -\frac{\sin\delta}{(2 - \lambda)\cos\delta 
+ y} \  , 
\eqno(34)
$$
and $\delta_{Li}$ and $\delta_{Ri}$ are unphysical phases and we may take 
$\delta_{Li} = \delta_{Ri} = 0$. The four and five 
parameters in $\widetilde{\overline{M}}_u$ and $\widetilde{\overline{M}}_d$, 
respectively, together with the phase parameter $\widetilde{\phi}$, 
are sufficient to fix the ten observable quantities, $D_u$, $D_d$ and $V$. 

We show the numerical results of 
$(\widetilde{M}_u, \ \widetilde{M}_d)$ 
without the approximation (25), for the case $\delta= \pi/2$:
$$
\begin{array}{l}
\widetilde{M}_u = 170 {\rm GeV} \left(\begin{array}{ccc}
0 & 0.000012 & 0 \\
0.0037 & 0 & -0.0669 \\
0 & 0 & 1 \\
\end{array} \right) \  , \\
\widetilde{M}_d = 2.52 {\rm GeV} \left(\begin{array}{ccc}
0 & 0.0081 & 0 \\
0.0074\ e^{i\widetilde{\phi}} & 0 & 0.0666\ e^{i\widetilde{\phi}} \\
0 & 0.5117 & 1 \\
\end{array} \right) \  , \\
\end{array} \eqno(35)
$$
$\widetilde{\phi} = -38.0^\circ$, where we have used the CKM parameter 
values [10], $|V_{us}| = 0.2205$,  $|V_{cb}| = 0.0041$, and 
$|V_{ub}/V_{cb}| = 0.08$, and the running quark mass values [11] at 
$\mu = m_Z$,  $m_u = 0.00222$ GeV, $m_d = 0.00442$ GeV,  
$m_s = 0.0847$  GeV, $m_c = 0.661$ GeV, 
$\ m_b = 2.996$ GeV and $m_t = 180$ GeV. 
Each value in (35), except for $\widetilde{M}_{u23}$ and $\widetilde{\phi}$, 
is insensitive to the value of the $CP$ violating phase $\delta$.

Although so far we have investigated the NNI form $(\widetilde{M}_u, \ 
\widetilde{M}_d)$ starting from the quark-family basis in which $M_u$ is 
diagonal, we can also discuss the NNI form starting from a quark-family 
basis in which $M_d$ is diagonal. Then, we get a similar result 
$\widetilde{M}_{d32} = 0$. This means that  
$\widetilde{M}_{d32}$ also depends on the phase convention of $V$. 

In conclusion, we have expressed the quark mass matrix $(\widetilde{M}_u, \ 
\widetilde{M}_d)$ with the NNI form in terms of the observable quantities 
as (33) with (31), (32) and (34). 
We have found that $\widetilde{M}_{u32}$ and $\widetilde{M}_{d32}$ depend 
on the phase convention of $V$ and we can always $\widetilde{M}_{u32} = 0$ 
without changing any physical situation. 
Every $|\widetilde{M}_{dij}|$ is insensitive to the value of the $CP$ 
violating phase parameter $\delta$, so that it can be fixed by the 
observed down-quark masses and CKM matrix parameters $|V_{us}|$, 
$|V_{cb}|$ and $|V_{ub}|$ only. 

By the way, we know that the rephasing of physical quark fields $(u_L, d_L)$, 
i.e., 
$$
\begin{array}{l} 
u_L \rightarrow u'_L = P(\delta^u) u_L \ , \\
d_L \rightarrow d'_L = P(\delta^d) d_L \ , \\
\end{array} \eqno(36)
$$
does not change the mass matrices $(H_u, H_d)$ for the fields 
$(u_L^0, d_L^0)$, although the ``re-basing" of quark fields 
$(u_L^0, d_L^0)$, i.e., 
$$
\begin{array}{l} 
u_L^0 \rightarrow u^{0\prime}_L = A u_L^0 \ , \\
d_L^0 \rightarrow d^{0\prime}_L = A d_L^0 \ , \\
\end{array} \eqno(37)
$$
changes the mass matrices $(H_u, H_d)$ 
as 
$$
\begin{array}{l} 
H_u \rightarrow H'_u = A H_u A^\dagger \ , \\
H_d \rightarrow H'_d = A H_d A^\dagger \ , \\
\end{array} \eqno(38)
$$
where $A$ is an arbitrary unitary matrix. 
Nevertheless, why have our results 
$(\widetilde{H}_u, \widetilde{H}_d)$ depended 
on the ``rephasing of $V$"?

Exactly speaking, our rephasing $V\rightarrow V'=P(\alpha) V$ in (3) does 
not mean the rephasing of physical quark fields $(u_L, d_L)$. 
The transformation (36) and (37) correspond to 
$$ 
\begin{array}{l} 
U_L^u \rightarrow U_L^{u\prime} = P(\delta^u) U_L^u A^\dagger \ , \\
U_L^d \rightarrow U_L^{d\prime} = P(\delta^d) U_L^d A^\dagger \ . \\
\end{array} \eqno(39)
$$
For the case of $(\widehat{H}_u, \widehat{H}_d)$ defined by (3), 
$U_L^u$ and $U_L^d$ are given by $U_L^u={\bf 1}$ and $U_L^d=V^\dagger$. 
The ``rephasing of $V$", $V\rightarrow V'=P(\alpha)V$, 
in (3) means that
$$ 
\begin{array}{lll} 
U_L^u={\bf 1} & \rightarrow & 
U_L^{u\prime} = P(\delta^u)  A^\dagger \ , \\
U_L^d=V^\dagger & \rightarrow & 
U_L^{d\prime} = P(\delta^d) V^\dagger P(\alpha) A^\dagger \ , \\
\end{array} \eqno(40)
$$
so that the phase matrix $P(\alpha)$ cannot be absorbed into the 
physical down-quark fields $d_L$.
Only when we take ``re-basing" $A=P(\alpha)$, the transformation 
(40) is expressed as 
$$ 
\begin{array}{lll} 
U_L^u={\bf 1} & \rightarrow & U_L^{u\prime} = P(\delta^u-\alpha) \ , \\
U_L^d=V^\dagger & \rightarrow & 
U_L^{d\prime} = P(\delta^d) V^\dagger \ , \\
\end{array} \eqno(41)
$$
where $P(\delta^u-\alpha)$ and $P(\delta^d)$ can be absorbed into 
the physical quark fields $u_L$ and $d_L$, respectively.
Thus, our ``rephasing of $V$" in (3) has tacitly been
accompanied with the ``re-basing" of quark fields $(u_L^0, d_L^0)$, 
so that our results $(\widetilde{M}_u, \widetilde{M}_d)$ have 
depended on the ``rephasing of $V$".
Note that such a problem does not appear when we discuss the 
observable quantities $D_u$, $D_d$ and $V$ starting from  a mass 
matrix $(M_u, M_d)$, but it appears only when we discuss the 
textures of $(M_u, M_d)$ starting from the observable quantities 
$D_u$, $D_d$ and $V$.

The present problem which was demonstrated for the NNI mass matrix 
also appears in the general study of 
$(M_u, M_d)$ when we want to express $(M_u, M_d)$  in terms 
of the observable quantities $D_u$, $D_d$ and $V$.
We would like to emphasize that we must take notice of 
the phase convention of $V$ when we investigate  [6]  
texture-zeros of $(M_u, M_d)$.

\vglue.3in

\centerline{\large\bf Acknowledgments}

The author would like to thank M.~Tanimoto, H.~Fusaoka, E.~Takasugi
 and S.~Wakaizumi for  helpful discussions. 
This work was supported by the Grant-in-Aid for Scientific Research, the 
Ministry of Education, Science and Culture, Japan (No.08640386). 

\vglue.2in

{\bf Note added}: After completion of this paper, I have recieved  an 
interesting paper by Takasugi [12], where he has found quark  mass matrix with
a symmetric form  by extending the present work. 

\vglue.3in
\newcounter{0000}
\centerline{\large\bf References}
\begin{list}
{[~\arabic{0000}~]}{\usecounter{0000}
\labelwidth=0.8cm\labelsep=.1cm\setlength{\leftmargin=0.7cm}
{\rightmargin=.2cm}}
\item N.~Cabibbo, Phys.~Rev.~Lett. {\bf 10}, 531 (1963); 
M.~Kobayashi and T.~Maskawa, Prog.~Theor.~Phys.~{\bf 49},  652 (1973).
\item For instance, Y.~Koide, H.~Fusaoaka and C.~Haba, Phys.~Rev. 
{\bf D46}, R4813 (1992); 
G.~B\'{e}langer, E.~Boridy, C.~Hamzaoui and G.~Jakimow, Phys.~Rev. 
{\bf D48}, 4275 (1993).
\item Y.~Koide, Phys.~Rev. {\bf D46}, 2121 (1992).

\item G.~C.~Branco and J.~I.~Silva-Marcos, Phys.~Lett. {\bf B331},
390 (1994); 
B.~Dutta and Nandi, Phys.~Lett. {\bf B366}, 281 (1996);
T.~Ito and M.~Tanimoto,  Phys.~Rev. {\bf D55}, 1509 (1997). 
\item K.~Harayama and N.~Okamura, Phys.~Lett. {\bf B387}, 617 (1996).
\item For instance, P.~Ramond, R.~G.~Roberts and G.~G.~Ross, 
Nucl.~Phys. {\bf B406}, 19 (1993).
\item H.~Fritzsch, Phys.~Lett. {\bf 73B}, 317 (1978); 
Nucl.~Phys. {\bf B155}, 182 (1979); 
L.~F.~Li, Phys.~Lett. {\bf 84B}, 461 (1979).
\item G.~C.~Branco, L.~Lavoura and F.~Mota, Phys.~Rev. {\bf D39}, 
3443 (1989).
\item L.~L.~Chau and W.-Y.~Keung,  Phys.~Rev.~Lett. {\bf 53}, 1802 (1984).
\item Particle data group, R.~M.~Barnet {\it et al}, Phys.~Rev. {\bf D54}, 
1 (1996).
\item H.~Fusaoka and Y.~Koide, in preparation.
\item E.~Takasugi, preprint OU-HET 264 (hep-ph/9705263) (1997).
\end{list}
\end{document}